
\input phyzzx

\def\prl{Phys. Rev. Lett. }

\overfullrule=0pt
\tolerance=5000
\overfullrule=0pt
\twelvepoint

\REF\WEE{X.G. Wen, Phys. Rev. {\bf B40}!1989)7378; Int. J. Mod. Phys.
{\bf B2}(1990)239.}
\REF\HR{ F.D.M. Haldane and E.H. Rezayi, Phys. Rev. {\bf B31} (85) 2529.}
\REF\HRN{ F.D.M. Haldane, Phys. Rev. Lett {\bf 55} (85) 2095.}
\REF\GP {{\it The Quantum Hall Effect},
Springer-Verlag, New York, Heidelberg,
1990. edited by R. Prange and S. Girvin.  2nd ed.}
\REF\ZHK {S.C. Zhang, T.H. Hansson, and S. Kivelson, Phys. Rev.
Lett. {\bf 62}(1989)82.}
\REF\RE {N. Read, Phys. Rev. Lett. {\bf 62}(1989)86.}
\REF\ROBERTO{R. Iengo and K. Lechner, Phys. Rep. {\bf 213}, 179 (1992).}
\REF\LID{Dingping Li,
{\it Hierarchical Wave Functions of Fractional Quantum
Hall Effect on the Torus}. SISSA/ISAS/2/92/EP.}
\REF\LAO{R.B. Laughlin, \prl {\bf 50}(1983)1395.}
\REF\HALD{F.D.M. Haldane, \prl {\bf 51}(1983)605.}
\REF\HALP{B.I. Halperin, Phys. Rev. Lett. {\bf 52}(1984)1583.}
\REF\REA { N. Read, Phys. Rev. Lett. {\bf 65}(1990)1502.}
\REF\WQ{X.G. Wen and Q. Niu, Phys. Rev. {\bf B41}(1990)9377.}
\REF\GREAD{G. Moore and N. Read, Nucl. Phys. {\bf 360B}(1991)362.}
\REF\BW {B. Blok, X.G. Wen, Phys. Rev. {\bf B42}(1990)8145; ibid. {\bf
43}(1991)8337.}
\REF\LIN{Dingping Li, Phys. Lett. {\bf A169}(1992)82;
{\it Anyons and Quantum Hall Effect on the Sphere},
SISSA/ISAS/129/EP.}
\REF\ILN{R. Iengo and K. Lechner, Nucl. phys. {\bf 365}(1991)551.}
\REF\CONWAY{J.H. Conway and N.J.A. Sloane, {\it Sphere Packings, Lattices and
Groups}. Springer-verlag, 1988.}
\REF\EIV{ T. Einarsson, Phys. Rev. Lett. {\bf 64} (90)1995.}
\REF\ASW {D.P. Arovas, J.R. Schriffer, and F. Wilczek, Phys. Rev.
Lett. {\bf 53}(1984)722.}
\REF\WI{{\it Fractional Statistics and  Anyon Superconductivity},
World Scientific, 1990. edited by F. Wilczek.}
\REF\BRA{X.G. Wen, E. Daggoto, and E. Fradkin, Phys. Rev. {\bf B42}(1990)6110.}
\REF\BRAA{Y. Hatsugai and M. Kohmoto, and Y.S. Wu,
Phys. Rev. {\bf B43}(1991)10761.}
\REF\ZHANG{S.C. Zhang, Int. J. Mod. Phys. {\bf B6}, 25 (1992).}
\REF\EIVV{ T. Einarsson, Mod. Phy. Lett. {\bf B5} (91)675.}
\REF\IL{R. Iengo and K. Lechner, Nucl. Phys. {\bf B346}(1990)551.}
\REF\KURT{K. Lechner, Phys. Lett. {\bf B273}(1991)463.}

\pubnum{SISSA/ISAS/58/92/EP}
\date{4/1992}
\titlepage
\title{Hierarchical Wave Functions and Fractional Statistics
in Fractional Quantum Hall Effect on the Torus}
\vglue-.25in
\author{Dingping Li}
\address{ International School for Advanced Studies,
I-34014 Trieste, Italy}
\bigskip
\abstract{One kind of hierarchical wave functions of Fractional Quantum
Hall Effect (FQHE) on the torus are constructed.
The multi-component nature of
anyon wave functions and the degeneracy of FQHE on the torus are very clear
reflected in this kind of wave functions. We also calculate the
braid statistics of the quasiparticles in FQHE on the torus and show they fit
to the picture of anyons interacting with magnetic field on the torus
obtained from braid group analysis.}
\endpage

\chapter{Introduction}

It is of great interest  to see that how FQHE
is realized on the surfaces of different topologies.
FQHE on the torus is particular interesting because torus provides
the simplest example of surface with nontrivial topology.
Recently  it has been emphasized that
a new kind of order called topological
order which appears in FQHE (Hall fluid)[\WEE],
the chiral spin fluid and the anyon superfluid.
FQHE on the torus at filling ${1\over m}$ with m being an integer has
$m$-fold center-mass degeneracy [\HR,\HRN].  This degeneracy actually is a
manifestation of the topological order.
Topological order describes
the global properties of the ground state which depend on the topology of
the surface and its low lying excitations.
The Field theory of such kind system is controlled by
topological field theory,
Chern-Simons theory , which is particularly relevant in the application to
the condensed matter systems.
The relation of Chern-Simons theory to FQHE has been investigated in
[\GP,\ZHK,\RE,\ROBERTO, etc.].
\par
In [\LID], a kind of  hierarchical wave functions
(for Laughlin [\LAO] wave function on the torus, see
[\HRN])
of FQHE on the torus is constructed
by generalizing the one [\REA] on the plane (the hierarchical construction
of FQHE state was first proposed in [\HALD,\HALP]).
The hierarchical state is
characterized by a generalized Abelian Chern-Simons theory.
Furthermore  the degeneracy is determined directly  from the wave functions,
which agrees with the prediction  in [\WQ,\GREAD].
\par
Another kind of hierarchical wave function of FQHE on the plane
have been constructed in [\BW] and analyzed by plasma analogue
(hierarchical wave function in [\BW] is based the wave function proposed in
[\HALP]; for the case on the sphere, see [\LIN]) and
this wave function has a very clear physical
picture,  the hierarchical condensation of the quasiparticles (holes of the
parents states).  To construct this kind of wave function, we  need
the wave functions of the condensed
quasiparticles in different hierarchy. Those wave
functions turn out to be multi-component on the torus. In this paper,
we shall  construct such kind of wave functions on the torus and
see what is the degeneracy of the wave functions.
FQHE at hierarchical filling on the torus also has been
investigated in [\ROBERTO,\ILN] in the context  of Chern-Simons theory.
\par
In the next section, the notations used in this paper are summarized
and some results in [\HR] are brief reviewed. In section $3$, the
hierarchical wave functions on the torus of the type as in [\BW]
are constructed following a simple
example. In section $4$, the fractional statistics of the quasiparticles in
FQHE on the torus is discussed. In section $4$,  we give some remarks about
the large gauge transformations and the modular transformations  of
the wave functions.

\chapter{Basic Notations and Haldane Wave Functions}

Following [\HR,\LID],
we consider a magnetic field with potential
${\bf A}=- B y $\^x, the wave function describing a electron in the
lowest Landau level has the form
$$\psi (x,y)=e^{-{ B y^2\over 2}}f(z) \, , \eqn\aaa$$
where $f(z)$ is the holomorphic function, and the units $e=1 \, , \hbar =1 $
are used.
It is better to use the lagrangian to analyse the symmetry of
the theory. The lagrangian of the electron in the magnetic field is
$$L=\sum_{i=1,2}{1\over 2}m {(v^i)}^2 +A^iv^i \, , \eqn\aab$$
where $L$ is invariant up to a total time derivative under the
translations. The corresponding Noether currents due to the translations are
$$t_x=m{\dot x}-By \, , t_y=m{\dot y}+Bx \, . \eqn\aac$$
The conjugate momenta are
$$p_x=m{\dot x}-By \,, p_y=m{\dot y} \, .\eqn\aad$$
So
$$t_x=p_x \, , t_y=p_y+Bx \, . \eqn\aae$$
They commute with Hamiltonian
$$H={1\over 2m} [{(p_x+By)}^2+{(p_y)}^2] \, ,\eqn\aaf$$
with the commutations $[x\, ,p_x]=i\, ,[y\, ,p_y]=i$ when the theory is
quantized.   we work on a torus  by identifying
$z \sim z+m+n \tau$ with $\tau =\tau_1 +i\tau_2$ and $\tau_2 \geq 0$.
The consistent boundary conditions imposed on the wave function of
the electron on this torus are
$$e^{it_x} \psi = e^{i\phi_1} \psi \, , e^{i\tau_1 t_x +i\tau_2 t_y}
\psi = e^{i\phi_2} \psi \, , \eqn\aag$$
with the condition $\tau_2 B=2\pi \Phi$, where $\Phi$
is an integer, which will insure that $e^{it_x}, e^{i\tau_1 t_x +i\tau_2 t_y}$
commute with each other for the consistence of the equation \aag.
By using the relation
$$e^{i\tau_1 t_x +i\tau_2 t_y}=e^{-i\tau_2 Bx^2 \over 2\tau_1}
e^{i\tau_1 p_x +i\tau_2 p_y}e^{i\tau_2 Bx^2 \over 2\tau_1}\, ,$$
\aag can be written as
$$f(z+1)=e^{i\phi_1} f(z) \, ,
f(z+\tau )=e^{i\phi_2}e^{-i\pi \Phi (2z+\tau)}f(z) \, . \eqn\aah$$
For the many-particle wave functions, the condition of the equation \aah
is imposed on every particle.
\par
The standard $\theta$ function is defined as
$$\theta (z|\tau ) =\sum _n \exp (\pi in^2 \tau +2\pi inz)\, ,
n\subset integer \, . \eqn\aai$$
More generally, the $\theta$ function \foot{In this paper, we only will use
$\theta$ function on one dimensional lattice. This kind of notation is
very helpful in the construction of the wave function.}
on the lattice [\CONWAY] is
$$\theta (z|e, \tau ) =\sum _{n_i} \exp (\pi i v^2 \tau +2\pi iv\cdot z)
\, , \eqn\aaj$$
where $v$ is a vector on a {\it l-dimension} lattice, $v=\sum_{i=1}^l n_i e_i$,
with $n_i$ being integers,  $e_i \cdot e_j =A_{ij}$ and $z=z_ie_i$.
The $\theta$ function in the equation \aai is a
special case of the $\theta$ function defined by \aaj with
$l=1, e_1 \cdot e_1 =1$.    Furthermore we define
$$\theta {a\brack b} (z|e, \tau ) =\sum _{n_i}
\exp (\pi i {(v+a)}^2 \tau +2\pi i(v+a)\cdot (z+b))\, , \eqn\aak$$
where $a, b$ are  arbitrary vectors on the lattice. Only the
positive matrix $A$ will be considered, which mean that $x_iA_{ij}x_j$
always is greater than zero when $x_i\not= 0$.
This requirement will insure that the $\theta$ function
in the equation \aaj is well defined.
The dual lattice $e^{\ast}_i$ is defined as
$$e_i^{\ast} \cdot e_j =\delta_{ij} \, , \eqn\aal$$
then we have $e^{\ast}_i\cdot e^{\ast}_j=A^{-1}_{i,j}$.
It can be verified that
$$\eqalign{& \theta {a\brack b} (z+e_i|e, \tau )=e^{2\pi ia\cdot e_i}
\theta {a\brack b} (z|e, \tau )\, , \cr & \theta {a\brack b}
(z+\tau e_i|e, \tau )=\exp {[-\pi i \tau e_i^2 -2\pi i e_i \cdot (z+b)]}
\theta {a\brack b} (z|e, \tau ) \, , \cr
& \theta {a\brack b} (z+e^{\ast}_i|e, \tau )=e^{2\pi ia\cdot e^{\ast}_i}
\theta {a\brack b} (z|e, \tau )\, , \cr & \theta {a\brack b}
(z+\tau e^{\ast}_i|e, \tau )=\exp {[-\pi i \tau {(e_i^{\ast})}^2
-2\pi i e^{\ast}_i \cdot (z+b)]}
\theta {a+e^{\ast}_i\brack b} (z|e, \tau ) \, , \cr}\eqn\aak$$
and
$$\theta {a+e_i\brack b+e^{\ast}_j} (z|e, \tau )=\exp (2\pi i a \cdot
e_j^{\ast}) \theta {a\brack b} (z|e, \tau ) \, . \eqn\aar$$
In {\it 1-dimension} lattice with $e_1 \cdot e_1 =1$, the $\theta$ function
is the one defined in the equation \aai. Moreover
$$\theta_3(z|\tau)=\theta {{1\over 2}\brack {1\over 2}}
(z|\tau )\, , \eqn\aal$$
is an odd function of $z$. And we have equations
$$\eqalign{& \theta_3(z+1|\tau)=e^{\pi i}
\theta_3(z| \tau )\, , \cr & \theta_3
(z+\tau | \tau )=\exp {[-\pi i \tau  -2\pi i \cdot (z+{1\over 2})]}
\theta_3(z|\tau ) \, . \cr} \eqn\aam$$
The Laughlin-Jastrow wave functions on the torus at the filling $1\over m$
($m$ is an odd positive integer) can be written as
$$\eqalign{&\Psi (z_i)=\exp (-{\pi \Phi \sum_i y^2_i \over
\tau_2})F(z_i)\, ,\cr & F(z_i)=\theta {a\brack b} (\sum_i z_ie|e,\tau)
\prod_{i<j} {[\theta_3(z_i-z_j|\tau)]}^m \, , \cr} \eqn\aan$$
where $\theta$  function is on {\it 1-dimension} lattice, $e^2=m$ , $i=1, 2
\ldots ,  N$ with $N$ being the number of the electron
and $a=a^{\ast}e^{\ast}, b=b^{\ast}e^{\ast}$. Thus
$$\eqalign{&F(z_i+1)={(-1)}^{N-1}e^{2\pi a^{\ast} }
F(z_i)\, , \cr & F(z_i+\tau)=\exp (-\pi (N-1)-2\pi i b^{\ast})
\exp [-i\pi mN(2z_i+\tau)] F(z_i)\, .\cr} \eqn\aao$$
Comparing to the equation \aah , we get
$$\Phi=mN, \phi_1=\pi  (\Phi +1)+2\pi n_1 +2\pi a^{\ast} ,
\phi_2=\pi (\Phi +1)+2\pi n_2 -2\pi b^{\ast} \, . \eqn\aap$$
 \aap has solutions
$$\eqalign{&a^{\ast}_i=a_0+i, \, \, ,
b^{\ast}=b_0  \, \, , i=0,1,\ldots , m-1  \, , \cr &
a_0={\phi_1 \over 2\pi }+{\Phi +1\over 2} \, \, ,
b_0=-{\phi_2 \over 2\pi }+{\Phi+1\over 2} \, , \cr}
 \eqn\aas$$
which  will give  $m$ orthogonal Laughlin-Jastrow
wave functions (other solutions are not independent on the solutions
given in \aas , which  can be seen from the equation \aar ).
So there is
$m$-fold center-mass degeneracy [\HR,\HRN] (see also the  discussion
in section $5$).

\chapter{Blok-Wen Hierarchical Wave Function on the Torus}

\section{An Example}

Then hierarchical FQHE in [\BW] describes a hierarchical condensation of holes
of the parent states.
The wave function can
be characterized by matrix $\Lambda$,
$$\Lambda =\pmatrix{p_1&+1&0&\ldots&0&0\cr
+1&-p_2&-1&0&\ldots&0\cr
0&-1&p_3&+1&0&\ldots\cr
\vdots&\vdots&\ddots&\ddots&\vdots&\vdots\cr
\vdots&\vdots&\ddots&\ddots&\vdots&\vdots\cr
0&\ldots&0&(-1)^{n-1}&(-1)^n p_{n-1}&(-1)^n\cr
0&0&\ldots&0&(-1)^n&(-1)^{n+1}p_n\cr} \, , \eqn\aat$$
where $p_1$ is a positive integer
(In the following discussion,
we will show that $p_i, i=2, 3, \ldots, n$ shall be  positive
even integers if we require that the wave function is well defined
on the torus).
$\Lambda$ describes a {\it n-level} hierarchical state.
The coordinates of the particles are expressed by  $z_{s,i}$. $z_{s,i}$
is the coordinate of the $i^{th}$ particle in level $s$, for example,
$z_{1,i}=z_i$ is the coordinate of the
$i^{th}$ electron.  We take a simple example,
$n=2$ hierarchical state, to demonstrate how to construct the hierarchical
wave function. In this case,  the wave function is supposed to be
$$\Psi(z_i)=\int \prod_\alpha dv_{2,\alpha} \sum_{l=0}^{p_1-1}
\Psi_1(z_i, z_{2,\alpha})_l
\Psi_2(z_{2,\alpha})_l \, , \eqn\aau$$
where $dv_{2,\alpha}=dz_{2,\alpha} d{\bar z}_{2,\alpha}$
which  are integrated
on the torus. $\Psi_1(z_i, z_{2,\alpha})_l$ in \aau are
the Laughlin wave functions of
electrons in the presence of the quasiparticles with the coordinates
$z_{2,\alpha}$ and $\Psi_2(z_{2,\alpha})_l$
is the Laughlin type wave function of
the quasiparticles. The index $l$ in  $\Psi_1$ is the degeneracy index
of the Laughlin wave functions with filling at $1\over p_1$. However
the index $l$ of  $\Psi_2$ is the component index of quasiparticle
wave function and it reflects the multi-component nature of anyon wave function
on the torus. In [\EIV], it is found that  free anyons
have  a multicomponent wave function on the torus
by using braid group analysis. However even  when anyons are exposed to
the magnetic field (the quasiparticles in FQHE interact with
the magnetic field),
by generalizing the results of [\EIV] to the case
of anyons interacting with magnetic field, the wave function  is
still found to be multicomponent (see the discussion in section $4$).
\par
The wave functions $\Psi_1$ are a Laughlin wave function with $N_2$
quasiparticles. Now we have the relation $p_1N_1+N_2=\Phi$   and
the wave functions are given by
$$\eqalign{\Psi_1(z_i, z_{2,\alpha})_l&=\exp (-{\pi \Phi
(\sum_i y^2_i+{1\over p_1} \sum_\alpha y^2_{2,\alpha}) \over
\tau_2})F_1(z_i)_l\, ,\cr  F_1(z_i)_l&=\theta {a_l\brack b}
(\sum_i z_ie+\sum_\alpha z_{2,\alpha} e^{\ast}|e,\tau)
\prod_{i<j} {[\theta_3(z_i-z_j|\tau)]}^{p_1} \cr
& \phantom{=}\times \prod_{i,\alpha} [\theta_3(z_i-z_{2,\alpha}|\tau)]
\prod_{\alpha < \beta} [\theta_3(z_{2,\alpha}-z_{2,\beta} |\tau)]^{1\over p_1}
\, , \cr} \eqn\aauu$$
where $e^2=p_1 \, , e^{\ast}={1\over e}$
and $a_l, b$ are still given by the equation \aas, e.g.
$a^{\ast}_l=a_0+l, b^{\ast}=b_0$.
As emphasized in [\BW], $\Psi_{1, l}$ needs to
be a normalized wave functions if we want to construct such
kind of hierarchical wave functions. Some may ask how we know $\Psi_{1,l}$ in
\aauu are the normalized wave functions?
The reason is that  everything is consistent in the end.
Another  reason is that $\Psi_{1,l}$ in
\aauu look like the normalized wave functions in Chern-Simons theory.
\par
Let us consider the wave function $\Psi_2$ now.  Firstly
$\Psi_2$ is a multicomponent (it has $p_1$ components)
wave function.  Secondly
when two quasiparticles are exchanged anti-clock,
the wave function in singular gauge
will give a phase $e^{i\theta}$ with $\theta =-{\pi \over p_1}$
(assuming inside the exchanging  path,
there are no other quasiparticles and we call $\theta$ as the statistical
parameter of the quasiparticles).
Thirdly it is Laughlin type wave function.
Under the magnetic translation of the quasiparticle, the wave function should
change up to a unitary transformation (the magnetic translation of
the electron is described by equation \aag;
for the case of the quasiparticle, see  section $4$).
The charge of the
quasiparticle of the Laughlin state
is $1\over p_1$.  Thus we find that
the wave function $\Psi_2$ shall be written as
(we write its complex conjugate),
$$\eqalign{&{\bar \Psi}_2 (z_{2, \alpha})_l=\exp (-{\pi
{\Phi \over p_1} \sum_\alpha y^2_{2,\alpha} \over
\tau_2})F_2(z_{2, \alpha})_l\, ,\cr
F_2(z_{2, \alpha})_l &=\theta {a_{2,l}\brack b_2}
(\sum_\alpha z_{2,\alpha} s_2|e_2, \tau)
\prod_{\alpha < \beta} [\theta_3(z_{2,\alpha}-z_{2,\beta} |\tau)]^
{{1\over p_1}+p_2}  \, , \cr} \eqn\bba$$
where $e^2_2=p_1(p_1p_2+1) \, , s^2_2=p_2+{1\over p_1}$
(the Laughlin type wave function of the quasiparticles on the torus
has also been discussed  in the context of Chern-Simons theory
[\ROBERTO,\ILN] and our construction of the wave function
agrees with them, but note that
here we work in {\bf anyon gauge}, which is different from [\ROBERTO,\ILN]).
{}From the form of $\Psi_2$, we can get the relation
$|q_2 \cdot \Phi| ={\Phi \over p+1}= N_2(p_2+{1\over p_1})$ where $q_2$ is
the charge of the quasiparticle and equals to
${1\over p_1}$ (assuming electron charge is $-1$).
Now we can write relations $p_1N_1+N_2=\Phi
\, ,{\Phi \over p+1}= N_2(p_2+{1\over p_1})$ as
$$\eqalign{&p_1N_1+N_2=\Phi \, ,\cr &
            N_1-p_2N_2=0 \, . \cr} \eqn\quasi$$
To fix the parameters $a_{2,l},b_2$,
we impose the condition that
$$\sum_{l=0}^{p_1-1}
\Psi_1(z_i, z_{2,\alpha})_l \Psi_2(z_{2,\alpha})_l$$
is periodic with the coordinates of the quasiparticles around
two nontrivial cycles of the torus in order that the integral in \aau
is well defined on the torus.
So we shall have
$$\eqalign{&\Psi_1(z_{2,\alpha}+1)_l\Psi_2(z_{2,\alpha}+1)_l=
\Psi_1(z_{2,\alpha})_l\Psi_2(z_{2,\alpha})_l \, ,\cr &
\Psi_1(z_{2,\alpha}+\tau)_l\Psi_2(z_{2,\alpha}+\tau)_l=
\Psi_1(z_{2,\alpha})_{l+1}\Psi_2(z_{2,\alpha})_{l+1} \, ,\cr}
\eqn\bbb$$
with $\Psi_{1,p_1}=\Psi_{1,0} \, , \Psi_{2,p_1}=\Psi_{2,0} $
where $\Psi_{1,p_1}=\Psi_1(z_i, z_{2,\alpha})_l
\, , \Psi_{2,p_1}=\Psi_2(z_{2,\alpha})_l$.
Then  we  get a set of solution of $a_{2,l}$ and $b_2$
$$\eqalign{&a_{2,l}=a^{\ast}_{2,l}{[p_1(p_1p_2+1)]}^{-{1\over 2}} \, ,
b_2=b^{\ast}_2 {[p_1(p_1p_2+1)]}^{-{1\over 2}} \, , \cr &
a^{\ast}_{2,l}=a_0+ l(p_1p_2+1)+
\lambda p_1 \, ,    b^{\ast}_2=b_0\, . \cr &
\lambda =0,1, \cdots, p_1p_2 \, , l=0, 1, \cdots , p_1-1 \, .
\cr} \eqn\bbc$$
The solution \bbc will give $p_1p_2+1$ independent wave functions $\Psi$,
which means that  the degeneracy of the electron ground states
is $p_1p_2+1$.
Now we write  the wave functions as
$$\Psi(z_i)_\lambda =\int \prod_\alpha dv_{2,\alpha} \sum_{l=0}^{p_1-1}
\Psi_1(z_i, z_{2,\alpha})_l
\Psi_2(z_{2,\alpha})_{l,\lambda} \, , \eqn\bbbn$$
where $\lambda$ is the index of the degeneracy
of the electron wave functions $\Psi$
and also is the index of the degeneracy of
the condensed quasiparticle wave functions
$\Psi_2$.  We also note that $p_2$ must be positive even integer, otherwise
the $\theta$ function in equation \bba will not be well defined.
\par

\section{GENERAL HIERARCHICAL WAVE FUNCTIONS}

For {\it n-level} hierarchical
wave functions,   we  define some useful parameters;
$d_m=|\det \Lambda (m)| \, , $
where matrix $\Lambda_{i,j}(m)=\Lambda_{i,j}
\, , 1\leq i,j \leq m $ is a $m\times m$ matrix with $d_0=1$, and
$$\eqalign{&e_m= (d_m \cdot d_{m-1})^{1\over 2} \,  , \cr
& s_m= {(d_m)^{1\over 2} \over  (d_{m-1})^{1\over 2}}  \, , \cr
& m=1,2, \cdots, n  \, .\cr} \eqn\aaz$$
Now the wave functions are
$$\eqalign{\Psi(z_i)_{\lambda_n}&=\int
\prod dv_{2,\alpha} \cdots dv_{n,\alpha}
\sum_{\lambda_1, \cdots, \lambda_{n-1}}
\Psi_1(z_{1,\alpha}, z_{2,\alpha})_{\lambda_1}
\Psi_2(z_{2,\alpha}, z_{3,\alpha} )_{\lambda_1, \lambda_2}   \cr
&\phantom{=} \cdots  \Psi_i(z_{i,\alpha},z_{i+1,\alpha}
)_{\lambda_{i-1}, \lambda_i} \cdots
\Psi_n(z_{n,\alpha})_{\lambda_{n-1}, \lambda_n}
\, , \cr} \eqn\bbd$$
where
$$\lambda_{i} =0,1, \cdots, d_i-1 \, .  \eqn\degen$$
Define
$$\eqalign{& {\tilde \Psi}_i(z_{i,\alpha},
z_{i+1,\alpha} )_{\lambda_{i-1}, \lambda_i}
={\bar  \Psi}_i(z_{i,\alpha},
z_{i+1,\alpha} )_{\lambda_{i-1}, \lambda_i}, i=odd \,\, integers \, , \cr &
{\tilde \Psi}_i(z_{i,\alpha},
z_{i+1,\alpha} )_{\lambda_{i-1}, \lambda_i}
={\Psi}_i(z_{i,\alpha},
z_{i+1,\alpha} )_{\lambda_{i-1}, \lambda_i}, i=even \, \, integers
\, . \cr} \eqn\bbde$$
Then when $1<i<n$,     we have
$$\eqalign{{\tilde \Psi}_i(z_{i,\alpha},
z_{i+1,\alpha} )_{\lambda_{i-1}, \lambda_i}
&=\exp (-{\pi \Phi
({1\over d_{i-1}}\sum_\alpha y^2_{i, \alpha}+
{1\over d_i} \sum_\alpha y^2_{i+1,\alpha}) \over
\tau_2}) \cr & \phantom{=} \times F_i(z_{i,\alpha},
z_{i+1,\alpha} )_{\lambda_{i-1}, \lambda_i} \, ,\cr
F_i(z_{i,\alpha}, z_{i+1,\alpha} )_{\lambda_{i-1}, \lambda_i}
&=\theta {a_{i, \lambda_{i-1},\lambda_i}  \brack b_i}
(\sum_{\alpha} z_{i, \alpha} s_i+
\sum_\alpha z_{i+1,\alpha} s^{\ast}_i|e_i,\tau) \times \cr &
\phantom{=}\prod_{\alpha<\beta}
{[\theta_3(z_{i, \alpha}- z_{i, \beta} |\tau)]}^{s^2_i}
\prod_{\alpha , \beta} [\theta_3(z_{i, \alpha}
-z_{i+1,\beta}|\tau)] \cr & \phantom{=} \times
\prod_{\alpha < \beta} [\theta_3(z_{i+1,\alpha}
-z_{i+1,\beta} |\tau)]^{{s^{\ast}_i}^2}
\, . \cr} \eqn\bbe$$
And the equation \quasi is generalized to
$$\sum_j \Lambda_{ij}N_j=\cases {\Phi, &if $i=1$; \cr 0, &otherwise. \cr}
\eqn\aav$$
where $N_j$ is the number of the condensed quasiparticles in level $i$.
Moreover the equation  \bbc becomes
$$\eqalign{&a_{i, \lambda_{i-1},\lambda_i}=
a^{\ast}_{i, \lambda_{i-1},\lambda_i}e^{\ast}_i \, ,
b_i=b^{\ast}_ie^{\ast}_i \, , \cr &
e^{\ast}_i={1 \over e_i}, s^{\ast}_i={1\over s_i} \, ,
\cr &a^{\ast}_{i, \lambda_{i-1},\lambda_i}=a_0
+ \lambda_{i-1} d_i +\lambda_i d_{i-1} \, ,
b^{\ast}_i=b_0 \, .\cr} \eqn\bbf$$
The condition in \bbf  needs to be satisfied in order that
$$\sum_{\lambda_{i-1}}
\Psi_{i-1}(z_{i-1,\alpha},z_{i,\alpha}
)_{\lambda_{i-2}, \lambda_{i-1}}
\Psi_i(z_{i,\alpha},z_{i+1,\alpha}
)_{\lambda_{i-1}, \lambda_i}$$
is periodic with the coordinates $z_{i, \alpha}$.
$\Psi_i(z_{i,\alpha},z_{i+1,\alpha}
)_{\lambda_{i-1}, \lambda_i}$  have  $d_{i-1}$ components
wave functions with $\lambda_{i-1}$ being the index of the components
of the wave functions
and have $d_i$ degeneracy with $\lambda_i$ being the index of the
degeneracy. These wave functions are the wave functions of the condensed
quasiparticles of the {\it i-level}  in anyon (singular) gauge.
If the electron charge is $-1$,
then the condensed quasiparticle in {\it i-level} has charge
$${{(-1)}^{i}\over d_{i-1}} \, , \eqn\char$$
and the statistics parameter $\theta$ of the condensed quasiparticle
(here we use the anti-clock exchange of
two quasiparticles and we will get a phase $e^{i\theta}$.
see also section $4$) is
$$ {(-1)}^{i-1}\pi {d_{i-2} \over d_{i-1}} \, . \eqn\stast$$
$\Psi_1(z_{1,\alpha}, z_{2,\alpha})_{\lambda_1}$ still are given by
the equation \aauu .  Finally,  we have
$$\eqalign{{\tilde \Psi}_n(z_{n,\alpha})_{\lambda_{n-1}, \lambda_n}
&=\exp (-{\pi \Phi
({1\over d_{n-1}}\sum_\alpha y^2_{n, \alpha}) \over
\tau_2})F_n(z_{n,\alpha})_{\lambda_{n-1}, \lambda_n} \, ,\cr
F_n(z_{n,\alpha})_{\lambda_{n-1}, \lambda_n}
&=\theta {a_{n, \lambda_{n-1},\lambda_n}  \brack b_n}
(\sum_{\alpha} z_{n, \alpha} s_n|e_n,\tau)
\prod_{\alpha<\beta}
{[\theta_3(z_{n, \alpha}- z_{n, \beta} |\tau)]}^{s^2_n} \, , \cr} \eqn\bbh$$
and $a_{n, \lambda_{n-1},\lambda_n}, b_n$ are still
given by the equation \bbf .
\par
Moreover $p_i, \, , i=2,3, \cdots, n$
shall be positive even integers in order that
the $\theta$ function appeared in the wave function be well defined.
\par
{}From equation \aav we can show that
the filling factor equals to $\nu={N_1 \over \Phi}=\Lambda^{-1}_{1,1}$,
where $N_1$ is the number of the electrons,
$$\nu ={1\over \displaystyle p_1+
{\strut 1\over \displaystyle p_2+
{\strut 1\over \displaystyle \cdots +
{\strut 1\over \displaystyle p_n}}}}  \, .\eqn\aax$$
Finally the degeneracy of the wave functions  $\Psi(z_i)_{\lambda_n}$ is
$$d_n=|\det \Lambda| \, , \eqn\deg$$
which actually is the denominator of the filling and
agrees with the prediction in the literature [\WQ,\GREAD].

\chapter{Fractional Statistics on the torus}

The quasiparticles in the FQHE satisfy fractional statistics
([\ASW]; see also  references in [\WI]).
In this section we will show that
the fractional statistics of the quasiparticles in the FQHE on the torus
can be directly calculated from the wave functions.
Hence this gives a concrete
example how the fractional statistics   can be realized on the torus.
In [\EIV], it has been proved that the fractional statistics of free anyons
on the torus is  consistent only
with multi-component wave functions (see also [\BRA,\BRAA]) and
claimed that the fractional quantum hall effect (FQHE)  fits to this picture.
It is worth to note that the quasiparticles in FQHE
interact with the magnetic field, so it needs to modify the results in [\EIV]
to the case that anyons interact with the magnetic field if one wants to
apply the results from  braid group
analysis to the quasiparticles of FQHE on the
torus.
We shall show that even if the anyons are exposed to the magnetic field,
the wave functions  will still need to be  multi-component.
We will calculate the braid statistics relation of
the quasiparticles in the Laughlin state  and compare it with the braid
statistics relation of  the anyons on the torus from
braid group analysis.

\section{Fractional Statistics of the Quasiparticles in FQHE on the torus}

The normalized wave functions of the simplest
Laughlin state with quasiparticles
in anyon (singular) gauge are given by the equation \aauu.
These wave functions will give a conjugate representation of the braid
statistics for the quasiparticles. This can be understood  as following; the
hierarchical construction of  wave functions are
$\Psi(z_i)=\int dw_{\alpha} \sum_l
\Psi_1(z_i,w_{\alpha})_l \Psi_2(w_{\alpha})_l$, where
$\Psi_2(w_{\alpha})_l$ are the wave functions of the quasiparticles,
$\Psi(z_i,w_{\alpha})_l$ are the normalized wave functions of the electrons
in the presence of the quasiparticles and both are in
singular gauge ($z_i$ are the coordinates of the electrons and
$w_{\alpha}$ are the coordinates of the quasiparticles).
{}From $\Psi_2(w_{\alpha})$, we can get the braid
statistics of the quasiparticles
and from $\Psi_1(z_i,w_{\alpha})$ we can get the complex conjugate
representation of the braid statistics of the quasiparticles.
Thus $\sum_l\Psi_1(z_i,w_{\alpha})_l \Psi_2(w_{\alpha})_l$
will give a trivial identity representation of the braid statistics
which is needed for the well
defined integration with the coordinates of the quasiparticles
on the torus.
What we discussed in the last section is
the case that the  wave functions of the
quasiparticles are Laughlin type and
the hierarchical state is obtained.
It is very natural to suggest
that even the  wave functions of the
quasiparticles are not  Laughlin type (the quasiparticles
are not condensed and may not have  Laughlin type wave function),
$\sum_l\Psi_1(z_i,w_{\alpha})_l \Psi_2(w_{\alpha})_l$
still give a trivial identity representation of the braid statistics
of the quasiparticles.
Thus  we can actually read out all braid statistics
relation of the quasiparticles from $\Psi(z_i,w_{\alpha})_l$
(or the wave functions given by \aauu) [\EIV]
even if we do not know the form of the wave
functions $\Psi(w_{\alpha})_l$ (if the quasiparticles are condensed and
have the Laughlin type wave functions, then $\Psi(w_{\alpha})_l$
are given by \bba).
Now we shall  demonstrate how to calculate the braid relation of
the fractional statistics of
the quasiparticles from $\Psi(z_i,w_{\alpha})_l$.
\par
The generators of the braid group are
$$\tau_i,\rho_i,\sigma_k;\, i=1,\cdots,n_q;\,k=1,\cdots,n_q-1 \, , \eqn\xaad$$
where $i$ is the index of anyon (quasiparticle in  FQHE)
and $n_q$ is the number of anyons.
The generators $\sigma_k$ are the anti-clockwise exchanges of anyons
$k$ and $k+1$ (we assume that there are
no particles in the  region of the exchange path).
The generators $\tau_i$ and $\rho_i$ are the {\bf magnetic} translation
operators of the particle $i$ along the fundamental non-contractible loops
of the torus (because of the presence of the magnetic field, we have
the magnetic translation symmetry instead of the translation symmetry).
\par
The  operators $i$, $\tau_i,\rho_i$
(acting on the wave function  $\Psi_2(w_{\alpha})_l$) are given by
$$\eqalign{&\tau_i=\exp(ip_{w_{i,1}})\, ,\cr&
\rho_i=\exp(i\tau_1p_{w_{i,1}}
+i\tau_2p_{w_{i,2}}-i2\pi {\Phi \over m}w_{i,1})\, ,\cr}\eqn\xaafnext$$
because the charge of the quasiparticles is $-{1\over m}$ of the one of
the electron,  where the filling of the Laughlin state
is taken as ${1\over m}$ and $w_{i,1}=Re(w_i)\, , w_{i,2}=Im(w_i)$.
\par
The magnetic translation operators of the
quasiparticle $i$, $\tau_i,\rho_i$ {\bf acting on}
the wave function  $\Psi_1(z_i,w_{\alpha})_l$ are given by
$$\eqalign{&\tau_i=\exp(ip_{w_{i,1}})\, ,\cr&
\rho_i=\exp(i\tau_1p_{w_{i,1}}
+i\tau_2p_{w_{i,2}}+i2\pi {\Phi \over m}w_{i,1})\, ,\cr}\eqn\xaaf$$
Actually $\tau_i,\rho_i,\sigma_k$ in \xaaf
will give the complex conjugate  representation of
the braiding operators of the quasiparticles in \xaafnext
on the torus, since
they act on  the wave functions $\Psi_1(z_i,w_{\alpha})_l$.
Assume $X$ is a braid operator and let $X$ act on
$\Psi_1(z_i,w_{\alpha})_l=\Psi_{1,l}$
and $\Psi_2(w_{\alpha})_l=\Psi_{2,l}$. Then we will have
$X\Psi_{1,l}=\Psi_{1,j}U_x(1)_{jl}$ where $X$
is an operator defined in \xaaf , and
$X\Psi_{2,l}=\Psi_{2,j}U_x(2)_{jl}$ where $X$
is an operator defined in \xaafnext .
$U_x(1)$ and $U_x(2)$ are unitary matrices.
Moreover, we have relation
$U_x(1)_{jl}=\bar U_x(2)_{jl}$, which will insure that
$\sum_l (X\Psi_{1,l}) (X\Psi_{2,l})=
\sum_l \Psi_{1,l}\Psi_{2,l}$ and this turns out  to mean that the function
$\sum_l \Psi_{1,l}\Psi_{2,l}$ is periodic with the coordinates
of the quasiparticles around two non-contractible loops
of the torus.
The operators $\tau_i,\rho_i,\sigma_k$ in \xaafnext
shall commute with
the Hamiltonian of the quasiparticles, which is given by
$$H=\sum_i^{n_q} {1\over 2M_q} [(p_{w_{i,1}}-{B\over m}w_{i,2})^2
+(p_{w_{i,2}})^2] \, .\eqn\saaf$$
We shall remark that, the lagrangian of
the quasi-particles are described by vortex (center coordinate) dynamics,
and the lagrangian of vortices does not contain any mass parameters [\ZHANG].
The Hilbert space of
the hamiltonian which we write above
shall be restricted to ground state.
The ground state of the above hamiltonian is the same as the
one  described by the vortex dynamic. To be rigorous, we shall
proceed from the vortex dynamic.
\par
Now we calculate the representation of the operators
$i$, $\tau_i,\rho_i$ {\bf acting on}
the wave function  $\Psi_1(z_i,w_{\alpha})_l$, which
are given in \xaaf.
For simplicity, we now choose $a_i=ie^{-1}, b=0, i=0,1,\cdots, m-1$ in \aauu.
Thus we have $\phi_1= \phi_2= \pi (\Phi-1)$
(any choices
$\phi_1, \phi_2$ will not affect the braid relations).
Now we denote the wave functions
$\Psi_1(z_i,w_{\alpha})_l$ as $\Psi_l$, a multicomponent column.
Furthermore we assume
$$\eqalign{&w_{1,1}<w_{2,1}<\cdots<w_{n_q,1} , \cr
&w_{1,2}<w_{2,2}<\cdots<w_{n_q,2} .  \cr} \eqn\xho$$
Then by applying $\sigma_i , \tau_i , \rho_i$ on $\Psi_l$, we have
$\sigma_i=e^{i\pi \over m}I_m$ with $I_m$ being
identity $m\times m$ matrix  and
$$\eqalign{&\tau_1=e^{\pi i (\Phi-1)\over m}\pmatrix{1&0&\cdots&0\cr
0&c&\cdots&0\cr \vdots&\vdots&\ddots&\vdots \cr 0&0&\cdots& c^{m-1}\cr}
\,. \cr &\rho_1=e^{-\pi i (\Phi-1)\over m}\pmatrix{0&\cdots&0&1\cr
1&\cdots&0&0\cr \vdots&\vdots&\ddots&\vdots \cr 0&\cdots&1&0\cr}\, ,\cr}
\eqn\xaah$$
where $c=e^{i2\pi \over m}$.
Other $\tau_i$ and $\rho_i$ are given by the
relation
$$\tau_{i+1}=e^{-2\pi i\over m}\tau_i, \rho_{i+1}
=e^{2\pi i\over m}\rho_i \, .\eqn\xaai$$

\section{Braid Group Analysis of the Fractional Statistics on the Torus}

The main relations of braid  statistics of free anyons are
[\EIV,\EIVV],
$$\eqalign{&\sigma_j\sigma_{j+1}\sigma_j=\sigma_{j+1}\sigma_j\sigma_{j+1}
\, ,\cr &
\tau_{j+1}=\sigma^{-1}_j\tau_j \sigma^{-1}_j\, ,
\rho_{j+1}=\sigma_j\rho_j \sigma_j \, , \cr &
\rho^{-1}_j\tau_{j+1}\sigma^{-2}_j\rho_j \sigma^2_j\tau^{-1}_{j+1}=\sigma^2_j
\, ,\cr & \sigma_1 \sigma_2 \cdots \sigma^2_{N-1}\cdots \sigma_2 \sigma_1=
\rho^{-1}_1\tau^{-1}_1\rho_1\tau_1 \, .\cr} \eqn\xaaj$$
If the anyons are exposed to the magnetic field, we just need to change the
last relation in \xaaj,
$$ \sigma_1 \sigma_2 \cdots \sigma^2_{N-1}\cdots \sigma_2 \sigma_1=
\rho^{-1}_1\tau^{-1}_1\rho_1\tau_1 e^{2\pi i q\Phi}
\, , \eqn\xaak$$
with $q$ being the charge of anyons and $\Phi$ being the magnetic flux out of
the torus.  The reason for the extra phase is that,
when we do the operation of  the left
equation \xaak, we get a closed curve with zero area which
can be deformed to the  operation of the right
equation [\EIV,\EIVV].  However, the curve needs to encompass the whole surface
during the deformation. So we get an extra phase (Aharonov
and Bohm phase)  because anyons now interact with the magnetic field.
\par
We can choose the base of the wave functions such that
$\sigma_i=e^{i\theta}I_M$ with $I_M$ as $M\times M$ identity
matrix [\EIV], then from the second and third equation in \xaaj, we get
$$\tau_i \rho_j=\rho_j \tau_i e^{2i\theta} \, . \eqn\yaaj$$
By taking the determinant of the equation \yaaj,
we have
$$\exp(2Mi\theta)=1 \, . \eqn\ya$$
If $\theta=\pi {r\over s}$ with $r$ and $s$ are coprime with each other,
so from \ya, we need $M=ns$ with $n$ being integer.
Furthermore from \xaak,  we have  another equation
$$e^{2 N\theta -2\pi q\Phi}=1 \, . \eqn\yab$$
which imply that ${Nr \over s}-q\Phi$ should be integer.
The braid statistics relation of the quasiparticles of FQHE in
above example  turn out to fit to the picture
from general braid group analysis.
Now $\theta ={\pi \over m}$ and the wave functions of
the quasiparticles are $m$ component, so \ya is fulfilled.
Furthermore in this example,  the equation \yab becomes
${n_q\over m}-{\Phi \over m}=integer$ with $n_q$ as the number of the
quasiparticles.
Because we have relation $mn_e+n_q=\Phi$ where $n_e$ is the electron
number,  so \yab is automatically satisfied.

\chapter{Remarks about the Modular Invariance and  Large Gauge Transformations}

The modular transformations and the large gauge transformations
of the wave functions in FQHE have been considered
in [\ROBERTO,\LID,\ILN]. How are about the wave functions constructed here?
Let us first discuss the modular transformations of the wave functions.
If we require the wave functions $\Psi (z_i)_{\lambda_n}$ are transformed
covariantly under the modular transformation, $z \to {z\over \gamma \tau +
\delta}$ and  $\tau \to {\alpha \tau +\beta \over \gamma \tau +\delta}$,
then we find  that $a_0,b_0$ need to equal to ${1\over 2}$.
The proving needs the modular
transformation of the quasiparticle wave functions
$\Psi_{m,n}$, where $m$ is the index of
the component and $n$ is the index of the degeneracy. Under the modular
transformations  $\tau \to \tau +1$ and  $z \to z$, the wave functions
$\Psi_{m,n}$ in the case of $a_0={1\over 2}\, ,b_0={1\over 2}$,
will be changed up a unitary phase. Under the modular
transformations  $\tau \to {- 1\over \tau} $ and  $z \to {z\over \tau}$,
$\Psi_{m,n} \to {\cal N} \sum_{m^{\prime}}\sum_{n^{\prime}}
f_1(m, m^{\prime}) f_2 (n, n^{\prime}) \Psi_{m^{\prime},n^{\prime}}$,
where ${\cal N}$ can depend on
the coordinates of the quasiparticles, but it
does not depend on $m^{\prime}, n^{\prime}$. These results have been
discussed in [\ROBERTO,\ILN,\KURT].
\par
Now we come to discuss
the large gauge transformations. Due to the nontrivial
topology of the torus, we have  gauge transformations like
$U_1=\exp ({-2\pi iy\over \tau_2})$ and  $U_2=\exp [{\pi (\tau {\bar z}-
{\bar \tau} z )\over \tau_2}]$. If we choose the magnetic potential as
$A_1=(-B y+{2\pi c_1\over \tau_2})$\^x, $ A_2={2\pi c_2\over \tau_2}$\^y
with $c_1, c_2$ being constant and $c=ic_1-c_2$, then under the gauge
transformations  generated by $U_1$ and $U_2$, we will have $c \to c+m+n\tau$.
When we take magnetic potential
$A_1=(-B y+{2\pi c_1\over \tau_2})$\^x, $ A_2={2\pi c_2\over \tau_2}$\^y,
the wave functions  $\Psi(z_i)_{\lambda_n}$
will depend on $c$. So we denote
the wave functions now as $\Psi(z_i, c)_{\lambda_n}$. The large gauge
transformations on the wave functions are defined as follows;
under the transformation $U_1$, $\Psi(z_i, c)_{\lambda_n} \to
\prod_i [U(z_i)_1]^{(-1)} G_1(c) \Psi(z_i, c+1)_{\lambda_n}$,  and
under the transformation $U_2$, $\Psi(z_i, c)_{\lambda_n} \to
\prod_i [U(z_i)_2]^{(-1)} G_2(c)\Psi(z_i, c+\tau)_{\lambda_n}$
(Under suitable choices of $G_1(c), G_2(c)$,  those operators actually form
an Heisenberg algebra [\ROBERTO,\ILN]).  We find that the wave
functions $\Psi(z_i, c)_{\lambda_n}$  change up to a phase under the gauge
transformation $U_1$ and change from one to  another under the
gauge transformation $U_2$.   Thus we can say that the degeneracy will
disappear by fixing the gauge of the magnetic field. This maybe offer
some reasons why this kind of the center coordinate degeneracy
of FQHE on the torus is {\bf quite} physically irrelevant [\HRN].
\par
The proving of the above statements is quite technical and complicated and
also because the discussion about the modular
transformations and large gauge transformation of the wave function
may not be physically interesting,
so we will not pursue it here and
just state the results.
The discussion of the modular transformations
and large gauge transformations  for another kind of hierarchical
wave functions on the torus can be found in [\LID].

\chapter{Summary and Conclusion}

In this paper, we have constructed one kind of hierarchical
wave functions on the torus.
In order to construct such wave functions on the torus, we
need first to have  the normalized quasiparticle  wave functions
on the torus. The parameters of the quasiparticle  wave functions
can be fixed  from the  requirement of
a well defined integration of the wave
functions on the torus by the coordinates of the quasiparticles.
Then the degeneracy of the wave function functions is obtained,
which shows how the topological order is manifested
in this kind of hierarchical  FQHE wave function on the torus and
agrees with the prediction in the literature.
We also calculate the braid matrix of the fractional statistics of
the FQHE quasiparticles on the torus and compare them to the general
results from braid group analysis.
Finally, the modular transformations and large gauge transformations of
the wave functions are briefly discussed, and it is found
that the parameters
can be fixed if we want the wave functions to be transformed covariantly
under the modular transformations,  and that the unique ground state
appears by fixing the gauge of the magnetic field.

\ack{The author would like to thank Prof. R. Iengo  for many stimulating
discussions and constant encouragements throughout this work.}

\refout
\end
\bye